\documentclass[twocolumn,english,aps,pra,superscriptaddress,amsmath,amssymb,floatfix,nofootinbib,longbibliography
]{revtex4-2}

\usepackage{amsthm}
\usepackage{amsfonts}
\usepackage{siunitx}
\usepackage{amsmath}
\usepackage{amssymb}
\usepackage{graphicx}
\usepackage{verbatim}
\usepackage[colorlinks]{hyperref}
\usepackage{tikz}
\usepackage{pgfplots}
\usepackage{adjustbox}
\usepackage{braket}
\usepackage{xcolor}
\usepackage{physics}
\usepackage{amssymb} 
\usepackage{graphicx}
\usepackage{dcolumn}
\usepackage{bm}
\usepackage{mathtools}
\usepackage{hyperref}
\usepackage{mathrsfs}
\usepackage{dashrule}
\usepackage{caption}
\usepackage{subcaption}
\usepackage{quantikz}

\usepackage[font=small,labelfont=bf,
   justification=justified,
   format=plain]{caption}

\definecolor{linkcolor}{RGB}{0,83,166}
\hypersetup{
  colorlinks = true,
  allcolors = {linkcolor}
}

\begin{document}

\title{Entanglement Negativity in Noisy Quantum Volume Sampling}

\author{Elijah Pelofske}
\email[]{epelofske@lanl.gov}
\affiliation{Quantum \& Condensed Matter Physics, Los Alamos National Laboratory, NM, USA}
\affiliation{Center for Quantum Computing, Los Alamos National Laboratory, Los Alamos, NM, USA}

\author{Stephan Eidenbenz}
\email[]{eidenben@lanl.gov}
\affiliation{Information Sciences, Los Alamos National Laboratory, NM, USA}
\affiliation{Center for Quantum Computing, Los Alamos National Laboratory, Los Alamos, NM, USA}

\begin{abstract}

The Quantum Volume protocol uses scrambling random circuits to benchmark NISQ computers. Quantum Volume is generally well-regarded as a benchmark for small, noisy, quantum computers because it requires the quantum computer to implement many non-local entangling gates within a square-shaped circuit, which incentivizes high qubit count, long qubit coherence times, and low error rates on all hardware gates. Quantum Volume circuits inherently produce high-entanglement states that are fragile to errors and decoherence. The Quantum Volume benchmark measures an observable called heavy-output-probability (HOP), where an HOP of $0.5$ corresponds to complete loss of coherence, and in the limit of system size an HOP $\approx 0.84$ for a fully coherent quantum processor. 
Here, we numerically study the tradeoff between depolarizing noise, entanglement as quantified by the bipartite negativity measure, and HOP in quantum volume circuits. 
Our results contextualize prior small scale quantum volume demonstrations on quantum computers and highlight that under depolarizing noise, due to finite system size effects heavy output probabilities can be greater than $0.5$ while the bipartite negativity entanglement has been destroyed. This implies, although improbable, that a NISQ computer could pass the Quantum Volume benchmark test threshold of $2/3$ while the underlying quantum computation has no global entanglement -- albeit only for small $n$.

\end{abstract}

\maketitle

\section{Introduction}\label{section:Introduction}

Quantum Volume (QV) is a well-designed and challenging noisy quantum computer benchmark algorithm~\cite{Cross_2019, aaronson2016complexitytheoreticfoundationsquantumsupremacy, Baldwin_2022} that tests the entire device. It requires a quantum processor to implement a large number of non-local entangling gates, and is defined as a square circuit whose depth, comprised of $d$ layers of Haar-random $\mathrm{SU}(4)$ (entangling) gates with random qubit index permutations in between each layer, is equal to the number of qubits $n$ in the circuit. It is a challenging benchmark for NISQ~\cite{Preskill_2018} computers as it requires low-error rates from every gate operation and component, and naturally rewards higher connectivity quantum computers. The Quantum Volume circuit inherently prepares a high-entanglement state that produces a Porter-Thomas-like distribution when sampled; highly entangled states are very sensitive to errors and qubit depolarization. If a quantum computer passes the Quantum Volume test for a circuit defined on $n$ qubits\footnote{Typically, this is reported as a Quantum Volume of $2^n$}, or passes similar random circuit scrambling algorithm protocols, that shows that the processor is able to prepare an entangled state of size $n$, which is an important quantum information processing demonstration~\cite{DeCross_2025}. 

A device passes a Quantum Volume test if it samples heavy-output probability (HOP) basis states in more than $2/3$ of all shots, where heavy-output probability states are defined such that they would be sampled, cumulatively, with a probability of about 84 percent on a noise-free device.  
However, here we demonstrate that, for small system sizes, the heavy-output-probability can be significantly higher than the entirely depolarized HOP sampling rate of $0.5$, even if the state on the device has lost all entanglement due to depolarizing noise. Interestingly, vice-versa, it can also be the case that the HOP observable is $0.5$, even while there is still some entanglement present in the state. We numerically probe this HOP/noise tradeoff, alongside the scaling of entanglement negativity as a function of quantum volume circuit size. Our principal question is how tightly correlated the HOP observable is with entanglement in noisy quantum systems. 

Our results indicate that -- in rare cases -- a device may pass a (low) QV test without holding any global entanglement (as measured by a bipartition), thus giving a ``false-positive'' result at least with respect to the spirit of the Quantum Volume test philosophy; on the flip side a device might fail a low quantum volume test despite retaining a small level of entanglement. These results do not put into question the validity and usefulness of the Quantum Volume test, but rather point out a previously unidentified mathematical property at the margins of the test regime. 

There have been many quantum volume demonstrations on various quantum computer architectures to date~\cite{jurcevic2021demonstration, Pelofske_2022, Pelofske_2024, Pino_2021, Moses_2023}. Our numerical results contextualize the level of global entanglement that exists, or does not exist, in depolarized noisy quantum volume sampling. 
Ref.~\cite{Li_2023} performed a related study on qudit random circuits, and Ref.~\cite{Zhang_2022} studied 1D random circuits. However, connecting noisy random circuit properties to the Quantum Volume benchmark~\cite{Cross_2019, aaronson2016complexitytheoreticfoundationsquantumsupremacy, Baldwin_2022} is a gap in the current literature.

\section{Methods}\label{section:methods}

\begin{figure*}[ht!]
    \centering
    \includegraphics[width=0.49\linewidth]{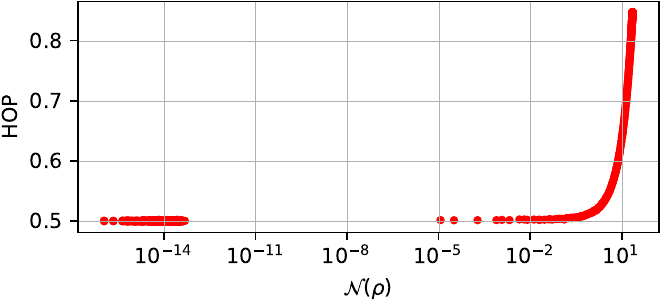}
    \includegraphics[width=0.49\linewidth]{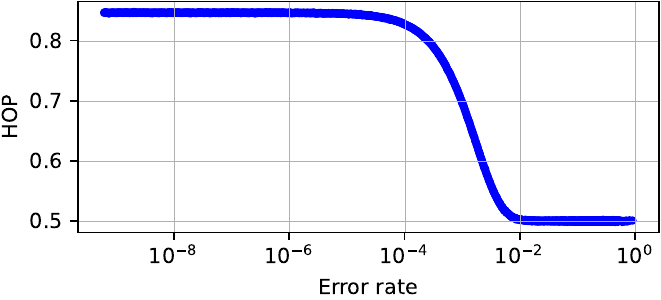}
    \includegraphics[width=0.49\linewidth]{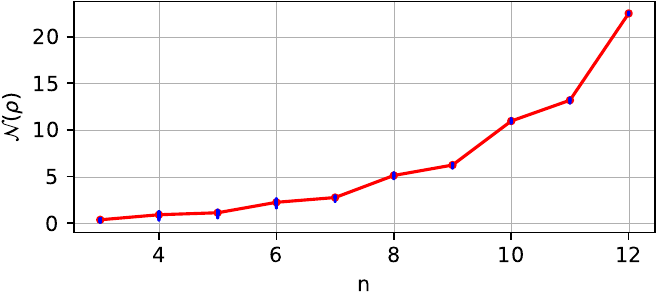}
    \includegraphics[width=0.49\linewidth]{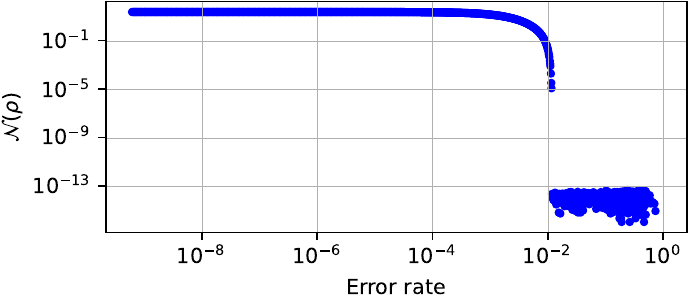}
    \caption{ \textbf{Tradeoff between depolarizing noise error rate, entanglement negativity, and heavy output probability, for one Quantum Volume circuit of $12$ qubits.} The \textbf{two right-most} subplots report HOP and $\mathcal{N}(\rho)$ as a function of the depolarizing noise error rate. The \textbf{top-left} subplot then shows HOP as a function of $\mathcal{N}(\rho)$, which notably shows that there can be some entanglement even when the heavy output probability is $0.5$, and importantly shows that there is a distinct separation between numerical floating point error accumulation and ``physical'' entanglement which has a minimum of $\mathcal{N}(\rho) \approx 10^{-5}$. Combined, the top left and bottom right plots show that there is a clear separation between numerical floating point error ($\approx 10^{-14}$) and genuine, if very low, entanglement -- there is a demarcating level of depolarizing noise which entirely removes all entanglement from the system. Lastly, the \textbf{lower-left} plot reports mean $\mathcal{N}(\rho)$ from an error-free statevector computation ($\lambda=0$) across the distribution of circuits as a function of $n$, which shows the expected growth in entanglement as the circuit sizes grow larger. 
    }
    \label{fig:HOP_negativity_and_error_rate}
\end{figure*}

In Quantum Volume, heavy output probability is defined as follows. Given an $n$-qubit quantum circuit $U$ that also has a depth of $n$ making it ``square'', the ideal probability distribution is 

\begin{equation}
    p_U(x) = \left|\langle x|U|0^n\rangle\right|^2, 
\end{equation}

where $x \in \{0, 1\}^n$, and we define the median probability as 

\begin{equation}
    m_U = \text{median} \{ p_U(x) : x \in \{0, 1\}^n \}. 
\end{equation}

The heavy output set of measured bit configurations is then defined as 

\begin{equation}
    H_U = \{ x \in \{0, 1\}^n : p_U (x) > m_U \}. 
\end{equation}

Then a quantum computer, measuring each QV circuit in the computational basis, samples a bitstring probability distribution $d_{QC}$. That quantum computer then has a measured heavy output probability of

\begin{equation}
    P_{\text{heavy}} = \sum_{x \in H_U} d_{QC}(x). 
\end{equation}

Asymptotically for large $n$-qubit QV circuits, the ideal heavy-output-probability~\cite{aaronson2016complexitytheoreticfoundationsquantumsupremacy} is 

\begin{equation}
P_{\text{heavy}} \approx \frac{1 + \ln(2)}{2} \approx 0.84. 
\end{equation}

The Quantum Volume test therefore requires a full classical statevector computation of the full probability distribution produced by a single circuit, which makes it intractable as a benchmark for very large coherent quantum computers, but it serves as a robust benchmark for NISQ computers. The QV benchmark is defined using the measured $P_{\text{heavy}}$ produced by a quantum computer for increasingly large QV circuit sizes $n$. If $P_{\text{heavy}}$ exceeds (for a given $n$), with high statistical confidence as determined by many random QV circuit instantiations, a threshold of $2/3$\cite{Cross_2019, Baldwin_2022, Mills_2021}, then that quantum computer has a quantum volume measure of $n$. 
A fully depolarized quantum computer will sample all measurements uniformly, resulting in a distribution produced by a fully mixed state $P_{\text{heavy}} = 0.5$. 

\begin{figure}[ht!]
    \centering
    \includegraphics[width=1.0\linewidth]{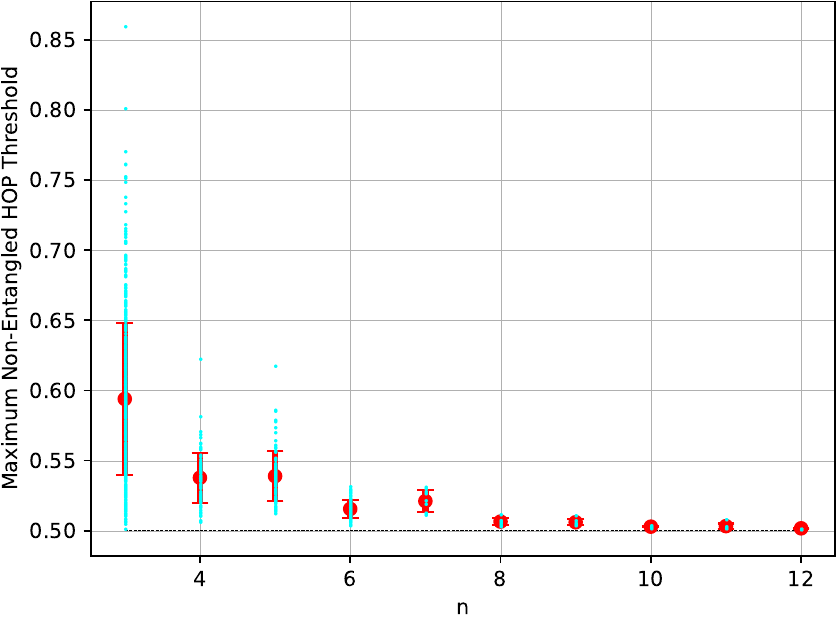}
    \caption{ \textbf{Non-entanglement HOP threshold: largest HOP distribution whose quantum state has no of physical (bipartite negativity) entanglement.} Each cyan dot corresponds to a single, random, quantum volume circuit, and each distribution at each $n$ has an error bar reported of one standard deviation (red). 
    }
    \label{fig:entanglement_threshold}
\end{figure}

Our goal here is to quantify the relationship between $P_{\text{heavy}}$, bipartite entanglement negativity, and depolarizing error. $P_{\text{heavy}}$ we will abbreviate as HOP. We use (bipartite) negativity to quantify entanglement in the depolarizing error quantum circuit simulations because it is a standard entanglement measure, has a clear definition of zero entanglement in depolarized quantum states, is easily computable given that a density matrix can be obtained, and is overall a standard entanglement measure that is widely used particularly for mixed quantum states~\cite{raza2026robustnegativityquantumtoclassicaltransition, ouyang2026logarithmicnegativitytypicallyequals, Lee_2013, Ruggiero_2016, Sang_2021, Calabrese_2013}. The negativity measure, given a density matrix $\rho$, is defined as

\begin{equation}
\mathcal{N}(\rho) = \frac{\vert \vert \rho^{T_A} \vert \vert_1 - 1 } {2}. 
\end{equation}

For each circuit, $\mathcal{N}(\rho)$ is computed using a single arbitrarily chosen bipartition of $|A|=\lfloor \frac{n}{2} \rfloor$. This means that there could be entanglement within a bi-partition while the overall $\mathcal{N}(\rho)$ equals zero, however, here we want to capture global entanglement in the QV circuits, therefore when $\mathcal{N}(\rho) = 0$ the depolarizing noise has destroyed coherent quantum entanglement. All numerical computations are carried out using Qiskit~\cite{javadiabhari2024quantumcomputingqiskit}, specifically the \texttt{qiskit-aer} classical simulator. The HOP probabilities are computed using shot noise ($10^{6}$ shots per circuit), thereby replicating how the real quantum volume protocol works on a quantum computer (e.g., the presence of statistical uncertainty due to shot noise). The entanglement negativity is computed using the full density matrix from the exact computation after the depolarizing quantum error channel has been applied to the circuit (specifically, every single and two qubit gate). No measurement or state preparation errors are modeled, only single and two qubit gate depolarizing error channels are applied to the quantum volume circuit once it has been decomposed and transpiled, with no circuit level optimization or simplification, into the universal gateset of \texttt{u3} and \texttt{cx}. The depolarizing error channel, implemented using \texttt{qiskit-aer}~\cite{javadiabhari2024quantumcomputingqiskit}, is defined as

\begin{equation}
E(\rho) = (1-\lambda)\rho + \lambda \Tr[\rho] \frac{\mathbb{I}}{2^n}, 
\end{equation}

where $\mathbb{I}$ is the identity matrix. The depolarizing error rate $\lambda$ is varied from $\approx 10^{-9}$ to $0.9$, in logarithmic spacing with reasonably high precision intervals. The noiseless ($\lambda=0$) case is also run. Each QV circuit is generated randomly, including random qubit index permutations. $500$ random QV circuits are generated and simulated for $n=3$, then $100$ circuits are generated and simulated for $n=4,5,6$, and for $n=7,8,9,10,11,12$ $10$ circuits are generated and simulated. Larger circuits were not simulated because of the high classical computational cost required to perform noisy density matrix simulations, and noiseless statevector simulations. Because of the random qubit permutation step, qubits can become non-interacting with the rest of the circuit by chance in the circuit creation process; therefore, when this occurs those circuits are discarded because it inherently would mean the circuit has no bipartite entanglement even in the noiseless case ($1$ occurred for $n=4$ and $15$ occurred for $n=3$). Depolarizing error is a simplified noise model, and is not a detailed quantum computer noise profile model, however it serves as a good representative error model for current quantum computers, whose qubits decohere the longer the computation lasts (e.g. in the absence of error correction). Numerically, $\mathcal{N}(\rho)$ will be non-zero due to floating point error accumulation, despite the ``physically'' meaningful level of entanglement being zero. We use a threshold of $\mathcal{N}(\rho) = 10^{-12}$ to determine an effective zero entanglement cutoff; non-zero negativity smaller than this is likely to be due to floating point error accumulation as opposed to ``physical'' entanglement in the circuit. The threshold of $10^{-12}$ is arbitrarily chosen; it is motivated by the numerical calculations run in this study, where we observe that $\mathcal{N}(\rho) \approx 10^{-14}$ to $\mathcal{N}(\rho) \approx 10^{-15}$ are computed even when $\lambda \approx 0.9$ (which is a very high depolarizing qubit error rate). Moreover, in the following section we present results that show this transition between HOP and $\mathcal{N}(\rho)$, which shows that there is a very clear delineation in the ``entangled'' vs the ``non-entangled'' regime as $\lambda$ is increased. 

\begin{figure*}[ht!]
    \centering
    \includegraphics[width=0.494\linewidth]{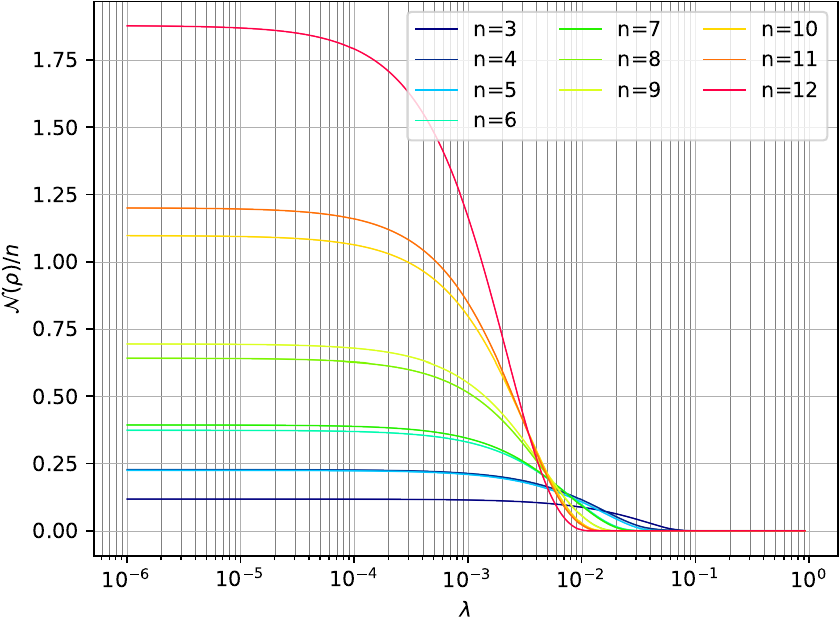}
    \includegraphics[width=0.494\linewidth]{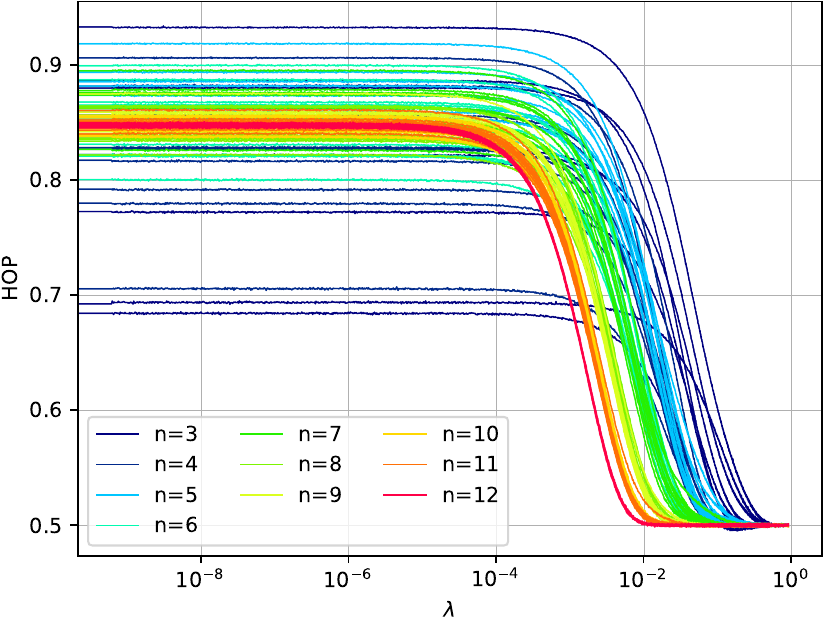}
    \caption{ \textbf{Aggregated negativity, normalized by system size, and HOP as a function of the depolarizing error rate $\lambda$.} The negativity is averaged over all random QV instances per $n$. The HOP distribution (right) is of $10$ QV circuits per $n$, where each individual HOP distribution is plotted for each circuit. At small system sizes, the HOP observable at high depolarizing error rates can dip slightly below $0.5$, which can be seen in one of the blue lines in the right sub-plot. 
    }
    \label{fig:error_rate_crossover}
\end{figure*}

\section{Results}\label{section:results}

In the context of random quantum circuit sampling protocols, in particular benchmark protocols such as linear cross-entropy or quantum volume, more entanglement corresponds to stronger anti-concentration which for quantum volume corresponds to higher heavy output probability~\cite{morvan2024phase, Boixo_2018, Bouland_2018, arute2019quantum, Zhang_2022}. These random circuits approximate Haar random states that then generate approximate Porter-Thomas statistics when measured in the computational basis, meaning these are highly entangled states~\cite{Harrow_2009}. The question we aim to examine is the exact threshold where all entanglement is lost when depolarizing noise is applied, and what HOP distributions occur at this fine-tuned noise threshold. 

Fig.~\ref{fig:HOP_negativity_and_error_rate} shows for a $n=12$ QV circuit the interplay between the depolarizing error rate, HOP, and entanglement negativity. This shows that there is a clear demarcating error rate that causes all entanglement, as measured by a bipartition, to be lost. Interestingly however, it is the case that there can be some detectable entanglement in the circuit while the HOP is very close to $0.5$. Moreover, we see as expected in the noiseless case that the entanglement grows consistently as $n$ increases. Next, we move to a system size analysis of the largest HOP that is measured while the entanglement is zero.

Fig.~\ref{fig:entanglement_threshold} reports the largest HOP value that was found from the numerical computations, for each circuit, that had no bipartite entanglement negativity. Here, the maximum HOP that is not globally entangled, as defined by negativity, is substantially larger than the expected HOP $\approx 0.5$ when $n$ is small. Although, notably the average non-entangled HOP does not exceed the $2/3$ HOP threshold that was originally proposed for the Quantum Volume protocol~\cite{Cross_2019}. Even at $n=12$, the average non-entangled HOP threshold is $0.50169$. One of the noteworthy observations from these results is that error-mitigation techniques such as zero noise extrapolation~\cite{giurgica2020digital, he2020zero} could be applied to these circuits with no entanglement in such a way that high HOP values (e.g., exceeding the $2/3$ threshold) could be found, while each individual circuit has no global entanglement. This motivates using ZNE for computations that are sufficiently large and where there is confidence that there is entanglement present in the quantum computation. Moreover, Fig.~\ref{fig:entanglement_threshold} shows that carefully, adversarially, chosen, rarely occurring, QV circuits could result in a QV benchmarking demonstration whose underlying quantum computations have no global entanglement. However, this is increasingly unlikely and hard to achieve as $n$ increases.

Fig.~\ref{fig:error_rate_crossover} plots the aggregated HOP and negativity measures as a function of depolarizing error rate. Interestingly this shows that the larger QV circuits are more fragile to depolarizing errors in terms of entanglement negativity; smaller error rates cause the negativity to go to zero as compared to $n=3,4$ QV circuits.

\section{Discussion and Conclusion}\label{section:conclusion}

We have numerically shown that quantum volume circuits, simulated under depolarizing error, can i) have some non-zero level of entanglement despite the heavy-output-probability being $0.5$ or ii) have a high heavy-output-probability, despite having no entanglement. This is an important caveat to quantum volume benchmarking for the NISQ-device performance community to be aware of, particularly for small quantum volumes, and high error rate devices~\cite{jaeger2024modelingquantumvolumeusing, Pelofske_2022, larose2022errormitigationincreaseseffective, Pelofske_2024}. These findings highlight a particular type of finite-system size effects in the high-entanglement random circuit scrambling protocol of Quantum Volume, which then informs real quantum computer hardware benchmarks. 

Importantly, despite this finite system size effect where the principal observable of quantum computer coherence, heavy output probability, can be substantially greater than the understood noisy decoherence, highly mixed state, threshold of $0.5$~\cite{Cross_2019, Baldwin_2022, Mills_2021}, it is the case that the arbitrarily chosen $2/3$ threshold~\cite{Cross_2019} can not be surpassed, on average for an ensemble of randomly produced QV circuits, if there is zero entanglement even for $n=3$ (see Fig.~\ref{fig:entanglement_threshold}). Single QV circuits can surpass this $2/3$ threshold, but on average the random circuit ensemble at $n=3$ has a mean non-entangled HOP of $\approx 0.59$. This bolsters existing quantum volume device benchmark data results, in that even for small QV circuits, for an ensemble of randomly produced QV circuits, there is a non-zero level of entanglement negativity produced by the quantum computation.

Systematic numerical studies of how depolarizing noise affects entanglement are important for a thorough characterization of these high-entanglement random circuit scrambling protocols such as Quantum Volume. This could be particularly important for the subset of these circuits recently described as \emph{peaked} quantum circuits~\cite{aaronson2024verifiablequantumadvantagepeaked}, because the level of entanglement correlates with how classically simulable these types of random circuit sampling protocols are.

%%%%%%%%%%%%%%%%%%%%%%%%%%%%%%%%%%%%%%%%%%
\section*{Acknowledgments}\label{sec:acknowledgments}
%%%%%%%%%%%%%%%%%%%%%%%%%%%%%%%%%%%%%%%%%%
This work was supported by the U.S. Department of Energy through the Los Alamos National Laboratory. Los Alamos National Laboratory is operated by Triad National Security, LLC, for the National Nuclear Security Administration of U.S. Department of Energy (Contract No. 89233218CNA000001), and by the NNSA's Advanced Simulation and Computing Beyond Moore's Law Program at Los Alamos National Laboratory. This research used resources provided by the Los Alamos National Laboratory Institutional Computing Program, and resources provided by the Darwin testbed at Los Alamos National Laboratory (LANL) which is funded by the Computational Systems and Software Environments subprogram of LANL's Advanced Simulation and Computing program (NNSA/DOE). LA-UR-26-27148

\bibliographystyle{apsrev4-2-titles}
\bibliography{references}
\end{document}